\documentclass[preprint]{aastex}
\received{}
\revised{}
\accepted{}
\shorttitle{Size and Shape of Voids}
\shortauthors{Schmidt, Ryden, \& Melott}
\begin{document}
\title{The Size and Shape of Voids in Three-Dimensional Galaxy Surveys} 
\author{Jason D. Schmidt }
\affil{Department of Physics, The Ohio State University,
Columbus, OH 43210}
\email{jd.schmidt@juno.com}
\author{Barbara S. Ryden }
\affil{Department of Astronomy, The Ohio State University,
Columbus, OH 43210}
\email{ryden@astronomy.ohio-state.edu}
\and
\author{Adrian L. Melott }
\affil{Department of Physics \& Astronomy, University of Kansas,
Lawrence, KS 66045}
\email{melott@kusmos.phsx.ukans.edu }

\begin{abstract}

The sizes and shapes of voids in a galaxy survey depend
not only on the physics of structure formation, but also
on the sampling density of the survey and on the
algorithm used to define voids.
Using an N-body simulation with a $\tau$CDM power spectrum,
we study of the properties of voids in samples
with different number densities of galaxies,
both in redshift space and in real space.
When voids are defined as totally empty regions
of space, their characteristic volume is strongly
dependent on sampling density; when they are
defined as regions whose density is 0.2 times
the mean galaxy density, the dependence is less strong.
We compare two void-finding algorithms, one
in which voids are nonoverlapping spheres,
and one, based on the algorithm of Aikio \&
M\"ah\"onen, which does not predefine
the shape of a void. Regardless of the algorithm
chosen, the characteristic void size is larger
in redshift space than in real space, and is
larger for low sampling densities than for
high sampling densities. We
define an elongation statistic $Q$ which measures
the tendency of voids to be stretched or squashed
along the line of sight. Using this statistic,
we find that at sufficiently high sampling
densities (comparable to the
number density of $L > L_*$ galaxies), large
voids tend to be slightly elongated along
the line of sight in redshift space.

\end{abstract}


\section{Introduction}
When the positions of galaxies are mapped in redshift space, the nature
of the large-scale structure seen in the maps is ``frothy'' or ``bubbly''.
Voids -- regions in which there are few or no galaxies -- fill much of
the map. The galaxies exist mainly in thin filaments or sheets that lie
between the voids. Comparison of the observed pattern of voids with that
predicted by a given model of structure formation is potentially a very
powerful test for acceptance or rejection of that model. Unfortunately,
the differences between models are often subtle. Many models for
structure formation predict a bubbly galaxy distribution. In the
gravitational instability scenario, regions that are originally underdense
expand faster than the Hubble flow \citep{fg84,be85}.
As underdense regions expand, they become more
nearly spherical \citep{fu83,ic84,bg90}.
Whether voids in the actual universe are viewed as
isolated spherical structures or as a space-filling foam
depends on the density level at which voids are defined
\citep{gm86}. Van de Weygaert \& van Kampen (1993) did
extensive studies of voids in constrained Gaussian fields,
finding that voids become nearly spherical in their very underdense 
inner regions, while their boundary regions remain more irregular.

If structure grows via gravitational instability, then the size and
shape of voids depends on the initial power spectrum $P(k)$ for
the density fluctuations and on the density parameter $\Omega$.
In an open universe, void evolution stops when $\Omega \sim 0.5$;
thereafter, the void network simply expands along with the Hubble
flow \citep{rg91}. At a given epoch, the mean void
radius is proportional to the nonlinearity scale $k_{\rm nl}$
\citep{km92}. The full distribution of void sizes, however,
depends on the shape of the spectrum $P(k)$ \citep{me87,rm96}.

If we know with absolute accuracy the position of galaxies in
real space, we could use the spectrum of void sizes to place
constraints on the initial power spectrum $P(k)$. However, measuring
the distances to galaxies is difficult; it is much easier to
measure their redshifts. Consequently, practical studies of
the properties of voids must take place in redshift space rather than
real space. If all galaxies smoothly followed the Hubble flow,
with no peculiar velocities, and if the Hubble constant $H_0$
were truly constant with time, then the mapping between
real space and redshift space would be linear. Generally,
though, the Hubble constant changes with time, and structure
that is isotropic in real space becomes distorted
along the line of sight in redshift space, with the amount
of distortion increasing with redshift $z$ \citep{ap79}.
Potentially, distortions in the
shapes of voids in redshift space can provide an
estimate of the deceleration parameter $q_0$ \citep{ry95}.
However, the distortions that result
from cosmological effects become large only when $z \sim 1$.
In the nearby universe, where $z < 1$, the dominant contribution
to distortions from redshift space comes from the peculiar
velocity of galaxies. \citet{rm96} demonstrated, for
instance, that in two-dimensional simulations with $P(k) \propto k^n$,
the characteristic void size is larger in redshift
space than in real space. Redshift space distortions
can also, in principle, be harnessed to measure $\Omega_m$,
the density in nonrelativistic matter \citep{mc98}.
One goal of this paper is to extend the analysis of \citet{rm96}
to three-dimensional simulations with
more realistic CDM power spectra.

Analyzing the properties of voids first requires a definition
of what a void is. Some of the statistics used to describe
voids define a void as a region totally devoid of galaxies.
The void probability function (VPF) is one such statistic;
the VPF $P_0 (V)$ is the probability that a randomly positioned
sphere of volume $V$ contains no galaxies \citep{wh79}. However,
the random dilution of a point process (selecting only a fraction
of all the galaxies in the universe, for instance), although it
leaves the correlation function unchanged, strongly affects
VPF \citep{sh96}. Thus,
from a practical viewpoint, it makes more sense to define a
void which is underdense with respect to the average number
density of galaxies. \citet{lw94}, for instance,
use the underdense probability function (UPF). The UPF $P_f
(V)$ can be defined as the probability that a randomly positioned
sphere of volume $V$ contains a number density of galaxies
less than or equal to $f$ times the average number density of
galaxies in the entire survey. We will follow \citet{lw94}
in setting a void threshold of $f=0.2$, matching the
density contrast of the largest voids in the CfA redshift survey
\citep{vg91,vg94}. One advantage
of the UPF over the VPF is that it is relatively insensitive
to the sparseness with which the galaxies are sampled. In this
paper, we will explicitly compare the UPF and VPF for different
galaxy sampling densities.

In addition to statistical measurements such as the VPF and UPF,
algorithms exist that identify individual voids within a sample
\citep{kf91,km92,ry95,rm96,ep97,am98}.
In addition to providing a spectrum of void sizes, these void-detection
algorithms also enable us to specify the location and shape of
individual voids. In this paper, we first use an algorithm (an
extension of the two-dimensional algorithm of Ryden [1995])
which defines voids as non-overlapping underdense spheres.
Although this algorithm is conceptually simple, it has the
disadvantage of forcing voids to be spherical. Thus, we also
employ a more sophisticated algorithm, based on that of
Aikio \& M\"ah\"onen (1998; AM), which does not constrain voids
to be any particular shape. The original AM algorithm defines
voids as empty regions; our modification of the original defines
voids as underdense regions. For the voids found by the modified
AM algorithm, we introduce an elongation statistic $Q$ which measures
whether the voids are stretched or squashed along the line of sight
from the origin. Just as with the UPF and VPF, we also apply the
void-detection algorithms to different sampling densities of
the same survey. Our studies of void properties in numerical
simulations will permit a more effective interpretation and
understanding of voids in future three-dimensional redshift
surveys.

In section 2, we describe the N-body simulation analyzed
in this paper. In section 3, we examine the statistical
properties of voids, as given by the VPF and UPF. In section 4,
we apply void-detection algorithms to the simulations. Finally,
in section 5, we analyze the effects of peculiar velocities
and galaxy sampling density on void properties, and discuss
implications for future work.

\section{The Numerical Simulation}

The simulation used for testing in this paper was done
using a particle-mesh (PM) N-body simulation. The PM method
is quite fast, and with a mean particle density of one
per simulation cell, represents the maximum resolution that
can be achieved without introducing two-body scattering
that decouples the result from its initial conditions on small
scales \citep{km96,sm98}.
The simulation used here had 256$^3$ particles on
a 256$^3$ mesh. Initial
conditions were generated by Fast Fourier Transform with random
phase perturbations. Since behavior at high density peaks is
not of interest to us in our study of voids, the simulations
were begun with an RMS density of 0.25 at the resolution limit.
The simulations were evolved until the RMS overdensity inside a
randomly located sphere of radius $8 h^{-1} {\rm\,Mpc}$ was
$\sigma_8 = 1.05$.

A matter-dominated Friedman-Robertson-Walker background density was
assumed, with the cosmological constant $\Lambda$ set equal to zero.
Since nonlinear modes are filtered out, and the dynamical effect
of nonzero $\Omega_\Lambda$ is well-understood in perturbation
theory \citep{lr91}, we did not use nonzero values of
$\Lambda$. The value of $\Omega_m$, the dimensionless matter density,
was taken to be $\Omega_m = 1$. The box size was take to be 1536 Mpc
and the Hubble constant $h = H_0/100 {\rm\,km} {\rm\,s}^{-1} {\rm\,Mpc}^{-1}
= 0.75$. Thus, the box size in redshift space corresponds to
$1.152 \times 10^5 {\rm\,km} {\rm\,s}^{-1} = 0.384 c$. Formally,
in the simulation $h$ only sets an overall timescale; since both
the expansion rate and the particle velocities scale with $h$,
the redshift space appearance does not change with $h$, only its
overall scale.

The initial power spectrum we assume is a cold dark matter (CDM)
spectrum, in which the parameter $\Gamma$ determines the shape of the
spectrum. Smaller values of $\Gamma$ are favored today, because they
push the turnover in the slope of the power spectrum to large
scales, in better agreement with observations. To test our void-finding
algorithms, we use $\Gamma = 0.25$. Normally, $\Gamma$ is taken
to be $\Omega_0 h$, as the turnover scale is set by the horizon
at the end of the radiation-dominated era. However, we break this
assumed coupling and take $\Gamma$ as a free parameter descriptive
of the spectral shape. Thus, our model corresponds to what is
sometimes called $\tau$CDM.

The number density of mass points in the simulation, $n = 4.63 \times
10^{-3} {\rm\,Mpc}^{-3}$, corresponds to the number density of galaxies
with $L > 0.42 L_*$, assuming that the luminosity function of galaxies
is a Schechter function with slope $\alpha = -1.07$ and normalization
$\Phi_* = 6.75 \times 10^{-4} {\rm\,Mpc}^{-3}$ \citep{ee88}.
One purpose of this
paper is to investigate the dependence of void properties on the
sampling density of a galaxy survey. Thus, from the initial numerical
simulation, we have created three different samples, corresponding
to volume-limited surveys with different luminosity cutoffs. To create
our densest sample, we randomly selected a fraction $X = 0.82$ of
the initial mass points, in order to match the number density of
galaxies with $L > L_*/2$. For the next densest sample, we selected
a fraction $X = 0.31$ of the mass points, to match the number density
of galaxies with $L > L_*$. Finally, for our least dense sample,
we selected a fraction $X = 0.067$ of the mass points, to match
the density of galaxies with $L > 2 L_*$. Note that by randomly
selecting galaxies in this way, we are assuming that galaxies are
unbiased with respect to the mass distribution, and that luminosity
segregation does not exist. We wish to study only the effects
of different sampling densities, and not the more subtle effects
of bias.

For each sampling density, we create two mock galaxy surveys, one
without peculiar velocities (which we will call the ``real space''
survey) and one with peculiar velocities (the ``redshift space''
survey). The mock surveys are spheres with a radius of
480 Mpc, or $z=0.12$ in redshift units. Note that in a flat
$\Omega_m = 1$ universe, a galaxy with $L=L_*/2$ located at
$z = 0.12$ will have an apparent magnitude of $m_B \approx 19$,
assuming $M_B^* = -19.7 + 5 \log h$ \citep{ee88}.
For a typical galaxy color of $r'-B \approx -1$, this
roughly corresponds to the flux limit $r' = 17.7$ expected for
the galaxy redshift sample of the Sloan Digital Sky Survey \citep{we00}.
A slice through the $L > L_*$
``real space'' survey is shown in Figure 1a; a slice through
the $L > L_*$ ``redshift space'' survey is shown in Figure 1b.

\section{Statistics of Voids}

The void probability function, or VPF \citep{wh79} has
been a widely used statistic for measuring the characteristic
size of voids. The VPF $P_0 (V)$ is defined as the probability
that a randomly located sphere of volume $V$ contains no
galaxies. The VPF has been applied to numerical simulations
\citep{fg89,ee91,wc92,lw94,vg94,gb94,gb96,cb96,kn97}
and to redshift surveys \citep{fg89,ee91,vg91,vg94,gb96}.
The VPF for the simulation studied in this
paper is given in the left-hand panels of Figure 2. The volumes
plotted are in redshift units; to convert to physical units,
multiply by $(c/H_0)^3 = 6.4 \times 10^{10} {\rm\,Mpc}^3$. In each panel,
the solid line shows the VPF in real space, and the dotted line
shows the VPF in redshift space. The effect of peculiar velocity
distortions is to increase the void probability at a given
volume $V$. A comparison of the VPF for the different sampling
densities, however, vividly illustrates the very strong dependence
of the VPF on the mean interparticle spacing. To illustrate,
let's define a characteristic void volume $V_{\rm VPF}$ as the volume
for which $P_0 (V_{\rm VPF}) = 0.01$. For a Poisson distribution of
points, $P_0 (V) = \exp (-\nu V)$, where $\nu$ is the
mean number density, and thus $V_{\rm VPF} = 4.61 / \nu$. For our
$L > L_*/2$ sample, $V_{\rm VPF} = 4.0 \times 10^{-8} = 9.6/\nu$;
for our $L > L_*$ sample, $V_{\rm VPF} = 1.14 \times 10^{-7} = 10.3/\nu$;
and for our sparse $L > 2 L_*$ sample, $V_{\rm VPF} = 4.3 \times 10^{-7} =
8.3/\nu$. Thus, the voids in the simulation
are larger than those in a Poisson distribution of equal
$\nu$, but it is still approximately true that the characteristic
void size (defining voids as totally empty volumes) is proportional
to $\nu^{-1}$.

The dependence of characteristic void size on the sample density
$\nu$ is reduced if we use the underdense probability function
(UPF) to measure the statistics of voids. The UPF $P_f (V)$ is
defined as the probability that a randomly located sphere of
volume $V$ contains a number density of galaxies that is less
than $f \nu$, where $0 < f < 1$, and $\nu$ is the mean number
density of galaxies in the sample. (In a flux-limited survey,
we can generalize the UPF so that $P_f (V)$ is the probability
that a sphere of volume $V$ located at a distance $r$ from
the origin contains a number density of galaxies that is
less than $f \nu(r)$, where $\nu(r)$ is the mean number density
of detected galaxies at $r$.) Following previous work
\citep{vg91,wc92,rm96},
we will set our density threshold at $f = 0.2$.
The UPF for the simulation studied in this paper is
given in the right-hand column of Figure 2; in each panel,
the solid line shows the UPF in real space and the dotted
line shows the UPF in redshift space. Again, the effect
of peculiar velocity distortions is to increase the probability
of finding a void of volume $V$.
For volumes $V < 1/(f\nu)$, the UPF and the VPF are identical,
since such a small volume can only fall below the density
threshold if it contains no galaxies. In general, a sphere
of volume $V$ will be underdense if it contains at most
$M$ galaxies, where $M = {\rm\,int} ( f \nu V )$. When $M$ is small,
the UPF shows discreteness effects, visible as the sawtooth
pattern in Figure 2; the UPF jumps upward whenever $V$ is
an integral multiple of $1/(f\nu)$.

The characteristic void size $V_{\rm UPF}$ is defined as the volume
for which $P_{0.2} (V_{\rm UPF}) = 0.01$. (The choice of
$P_0 = 0.01$ as the defining probability is somewhat arbitrary;
however, because of the rapid decline of $P_0$ at large volumes,
the exact value of $P_0$ is not crucial.)
For our dense $L > L_*/2$
sample, $V_{\rm UPF} = 2.7 \times 10^{-7} = 64/\nu$; for the
$L > L_*$ sample, $V_{\rm UPF} = 4.2 \times 10^{-7} = 38/\nu$;
for the sparse $L > 2 L_*$ sample, $V_{\rm UPF} = 9.2 \times 10^{-7} =
17.7/\nu$. The inclusion of peculiar velocity distortions
increases $V_{\rm UPF}$ by a factor which ranges from $2.5$
for the densest sample to $1.4$ for the sparsest sample.
Although the dependence of the underdense void size
$V_{\rm UPF}$ on $\nu$ is less strong than that of $V_{\rm VPF}$, it is
not true that $V_{\rm UPF}$ is independent of $\nu$ for plausible
sampling densities. Thus, in a flux-limited survey, where 
the measured $\nu$
decreases with distance from the origin, the characteristic
void size $V_{\rm UPF}$ will increase with distance.

\section{Void-detection Algorithms}

The UPF gives a statistical measure of the number and size
of voids in a given galaxy distribution. Frequently, however,
it is useful to identify individual voids instead of simply
giving a statistical description. Many different algorithms
have been used to detect and identify individual voids
\citep{kf91,km92,ry95,rm96,ep97,am98}.
In this paper, we will investigate
two void-detection algorithms that are distinguished by
their ease of use and clarity of conception. The first
algorithm, based on that of \citet{ry95}, identifies voids
as nonoverlapping spheres; we will call this the ``sphere
algorithm''. The second algorithm, based on that of
Aikio and M\"aho\"nen (1998; AM) permits voids to be nonspherical;
we will call this the ``AM algorithm''.

Both void-detection algorithms start by defining a continuous
scalar field $D_f ({\vec x})$ within the galaxy survey. At any
location $\vec x$, $D_f$ is defined as the radius of the largest
sphere centered on $\vec x$ within which the average galaxy
density is equal to $f\nu$. To implement the ``sphere algorithm'',
first locate the global maximum of $D_f$ within the survey;
call the location of this maximum ${\vec x}_1$. This is
the center of the largest spherical void in the sample,
which has a radius $D_f ({\vec x}_1)$. To find the second
largest void, find the point ${\vec x}_2$ for which $D_f$
is maximized, subject to the constraint that
\begin{equation}
D_f ({\vec x}_2) + D_f ({\vec x}_1) \geq | {\vec x}_2 - {\vec x}_1 | \ .
\end{equation}
The point ${\vec x}_2$ is then the center of the second-largest
spherical void, which has radius $D_f ({\vec x}_2)$. In other
words, the second-largest void is the largest underdense sphere
that doesn't overlap the largest void. Additional voids
are found by an iterative process. The $N^{\rm th}$ largest
void is located at the position ${\vec x}_N$ for which
$D_f $ is maximized, subject to the constraint that
\begin{equation}
D_f ({\vec x}_N ) + D_f ({\vec x}_i) \geq | {\vec x}_N - {\vec x}_i | \ ,
\end{equation}
for $i = 1, 2, \dots, N-1$.

In practice, we compute the values of $D_f$ on a cartesian
grid superimposed on the galaxy distribution. For the simulation
used in this paper, the grid spacing we used was $\Delta x = 0.0012$
in redshift units ($\Delta x = 4.8 {\rm\,Mpc}$ in physical units).
This is 0.8 times the resolution of the
original numerical simulation. Using a grid very much finer than
that of the original simulation is pointless, since there is no
information on such small scales. The computed value of $D_f$ for
each grid point was forbidden to be larger than the distance
from the grid point to the sample boundary at z=0.12.
We then located the spherical
voids, using the algorithm outlined above, subject to the additional
constraint that the void centers lie on grid points. Since
the discreteness of the superimposed grid creates errors of
order $\Delta x$ in the location of void centers, we halt the
void-detection algorithm when $D_f = 2 \Delta x$.

To emphasize the difference between the properties of totally
empty voids and underdense voids, we implemented the sphere
algorithm twice, once with $f = 0$, and once with $f = 0.2$.
To illustrate the voids found with the sphere algorithm,
Figure 3 shows a slice through the underdense spheres found
in the $L>L_*$ sample, setting $f=0.2$. Figure 3a shows
the spherical voids found in real space (without peculiar
velocity distortions) and Figure 3b shows the spherical voids
found in redshift space (with peculiar velocity distortions).

To show the spectrum of void sizes found with the sphere
algorithm, Figure 4 plots $F_f (V)$, the fraction of the total
volume of the sample found in spherical voids with volume $\geq V$.
The right column of Figure 4 displays $F_{0.2}$, the fraction
of the total volume found in voids with underdensity $f=0.2$.
For comparison, the left column of Figure 4 gives $F_0$, the fraction of
the total volume found in totally empty voids. The solid line
in each panel gives the distribution in real space, and the
dotted line gives the distribution in redshift space.
The left column of Figure 4 demonstrates, as in the case of
the VPF, that the characteristic size of empty voids depends
strongly on the density $\nu$ of the sample. Because of this
undesired feature, we will only examine in detail the properties
of underdense voids.

A characteristic void size can be found, following the
practice of \citet{km92}, by computing
the volume-weighted mean void size,
\begin{equation}
\overline{V} = \sum_1^{N_V} V_i^2 / \sum_1^{N_V} V_i \ ,
\end{equation}
where $V_i$ is the volume of the $i^{\rm th}$ largest void,
and the total number of voids is $N_V$. Alternatively,
we can define a characteristic void size $V_X$ which
is the void volume such that $X$ percent of the
total volume in the sample is contained in voids of
size $V_X$ or bigger. The characteristic void sizes
$\overline V$ and $V_{20}$ are given in Table 1 for
underdense voids in samples of different number density,
with and without peculiar velocities.

The characteristic size of spherical voids (whether defined
as $\overline V$ or $V_{20}$) increases in going from
real space to redshift space. The more densely sampled
the survey, the greater the increase in void size. For
the $L > 2 L_*$ sample, $\overline V$ increases by
43\% in going from real space to redshift space, and
$V_{20}$ increases by 40\%. For the $L > L_*/2$ sample,
$\overline V$ increases by 88\% in going from real space
to redshift space, and $V_{20}$ by 155\%.

Although the sphere algorithm gives a rough estimate
of the spectrum of void sizes, the volumes of the voids
found by this algorithm will generally be underestimates
of the ``true'' void size. That is, if a sample contains
an empty region surrounded by a continuous, well-defined,
extremely overdense wall, the void found by the sphere
algorithm will be the largest sphere that can be inscribed
within the wall. The remaining empty space within the void
wall will then be iteratively filled with smaller and smaller
spheres. A more flexible algorithm -- one which doesn't
impose the artificial constraint that voids are spherical --
should give a more accurate measure of void size.

In addition to underestimating the size of voids, the
sphere algorithm gives no hint of the true void shape.
In two dimensions, \citet{ry95} and \citet{rm96}
estimated the shapes of voids by fitting ellipses to
the underdense region. Each void was then characterized
by an axis ratio $q$ and a position angle $\phi$ in addition
to its area $a$ and position $(x,y)$.
Extending this algorithm to three dimensions
by fitting ellipsoids to the underdense regions becomes a
computationally daunting task. Each ellipsoidal void must be
characterized by two axis ratios and three Euler angles in
addition to its volume $V$ and position $(x,y,z)$. The
introduction of additional parameters makes the search through
parameter space far more time-consuming. Thus, instead of
approximating the shape of voids by fitting ellipsoids to
them, we adopted the more flexible scheme of Aikio \&
Mah\"on\"en (1998; AM).

To implement the AM algorithm, we start with the field $D_f$
as defined on the cartesian grid that we have superimposed
on the galaxy survey. If the grid spacing is $\Delta x$, then
each grid point can be thought of as being at the center
of a cubical ``elementary cell'' of volume $(\Delta x)^3$.
We identify the local maxima of $D_f$ on the grid as being
those points which have values of $D_f$ greater than that
of their 26 closest neighbors (the 6 points at a distance
$\Delta x$, the 12 points at a distance $\sqrt{2} \Delta x$,
and the 8 points at a distance $\sqrt{3} \Delta x$). We
label the local maxima we find as $M_1$, $M_2$, $\dots$,
$M_N$, where $N$ is the total number of maxima located.
The AM algorithm assigns every elementary cell to a ``subvoid''
associated with some maximum $M_i$. To discover which
subvoid a particular elementary cell belongs to, a
``climbing algorithm'' is used. For a elementary cell $X$,
we compute the gradient in $D_f$ to each of the neighboring
cells. We then ``climb'' to elementary cell $X'$ for
which the gradient has the largest (positive) value. The
climbing continues from cell to cell until a local maximum
$M_i$ is reached. The cell $X$ (and every other cell along
the climbing route) is then assigned to the subvoid of
maximum $M_i$. In this way, every elementary cell is assigned
to a subvoid, and each maximum $M_i$ has an associated
subvoid which consists of at least one elementary cell.
Once every elementary cell is assigned to a subvoid, the
subvoids are joined together into larger voids. The
subvoids associated with maxima $M_i$ and $M_j$ are
members of the same void if the distance between
$M_i$ and $M_j$ is less than both $D_f (M_i)$ and
$D_f (M_j)$. Using this criterion, all the subvoids
are grouped into voids; some voids contain a single
subvoid, while others contain many subvoids linked
together in a friends-of-friends percolation.
The AM algorithm and its implementation is described in
more detail in the original paper by \citet{am98};
our modification is to use the underdensity field $D_f$,
where AM restricted themselves to the case $f=0$.

Figure 5a shows a slice through the underdense voids
in the $L > L_*$ sample, without the inclusion of
peculiar velocity distortions. The AM algorithm was used
with a density threshold $f=0.2$. In each panel of Figure 5,
an elementary cell is colored white if it belongs to
the same void as its 26 nearest neighboring cells;
it is colored black if one or more of its neighbors
belongs to a different void. Figure 5b shows a slice
through the underdense voids in the $L > L_*$ sample,
this time with the effects of peculiar velocities included.

The volume $V$ of an individual void found by the AM algorithm
can be found by simply adding together the volume $(\Delta x)^3$
of the elementary cells which it contains. The full spectrum
of void sizes in a particular galaxy sample is given by $F_f (V)$,
the fraction of the total volume of the sample found in voids
with volume $\geq V$. The left column of Figure 6 shows $F_0$,
the distribution of void volumes using the original algorithm
of AM, in which $f = 0$. The right column of Figure 6 displays
$F_{0.2}$, using our usual underdense criterion, $f=0.2$.
Some properties of the AM voids are the similar to those of
the spherical voids found earlier. For instance, the left
column of Figure 6 shows that the characteristic size of
empty AM voids (just like those of spherical voids) is
strongly dependent on the density $\nu$ of the sample.
Also, peculiar velocities increase the characteristic
size of AM voids as well as of spherical voids. Using
the AM algorithm with $f = 0.2$, the volume-weighted
mean void size $\overline V$ and the characteristic
size $V_{20}$ are given in Table 1. For the sparsely
sampled survey ($L > 2L_*$), $\overline V$ increases
by 58\% in going from real space to redshift space,
and $V_{20}$ increases by 66\%. For the densely sampled
survey ($L > L_*/2$), $\overline V$ increases by 79\% in
going from real space to redshift space, and $V_{20}$
increases by 82\%. In addition, AM voids as well as
spherical voids are larger at lower sampling densities.
Given a density threshold of $f=0.2$, the characteristic
void size (either $\overline V$ or $V_{20}$) for the
sparsely sampled survey ($L > 2L_*$) is 2.5 times that
of the densely sampled survey ($L > L_*/2$).

One important difference between the voids found by the
AM algorithm and those found by the sphere algorithm is
that the AM voids are larger. At all sampling densities,
it is found that $\overline{V} ({\rm\,AM}) / \overline{V}
({\rm\,sphere}) \sim 2.7$. The ratio $V_{20} ({\rm\,AM}) /
V_{20} ({\rm\,sphere})$ ranges from $\sim 6$ for the
$L>2 L_*$ survey through $\sim 8$ for the $L > L_*$ survey
to $\sim 14$ for the $L>L_*/2$. Thus, the answer to the
seemingly innocuous question, ``How large is a typical void?''
depends not only on the sampling density of the galaxy survey
and on whether the survey is done in real space or in redshift
space, but also on the void-finding algorithm used and
on the definition adopted for the typical void size.

Another important difference between the voids found by
the AM algorithm and those found by the sphere algorithm
is that the AM voids are not compelled to be spherical.
Hence the shapes of the AM voids can be used as a measure
of the shape of voids in the galaxy distribution. It is
of particular interest to discover whether voids are
distorted along the line of sight from the observer
at the origin. At relatively small redshifts ($z \lesssim 1$),
the dominant source of distortion in redshift maps is
the peculiar velocities of galaxies. In examining
two-dimensional simulations with power spectra $P
\propto k^n$, \citet{rm96} found only a mild
tendency for voids to be distorted or compressed along
the line of sight. With an $n=2$ power spectrum, the largest voids
were slightly elongated along the line of sight; with an
$n=0$ spectrum, voids were slightly compressed along the line
of sight.

To find whether the AM voids are preferentially elongated
or compressed along the line of sight, we begin by computing
the moments of the voids. If a void contains $N$ elementary
cells, with the center of the $i^{\rm th}$ cell at $(x_i, y_i, z_i)$,
the moments of the void can be computed as
\begin{equation}
\overline{ x^\alpha y^\beta z^\gamma } = {1 \over N} \sum_{i=1}^N
x_i^\alpha y_i^\beta z_i^\gamma \ .
\end{equation}
The ``center of mass'' of the void, weighting all elementary
cells equally, is at $(\overline{x}, \overline{y}, \overline{z})$.
For a given void, we can create a new coordinate system, with
its origin at $(\overline{x}, \overline{y}, \overline{z})$,
with its $x'$ axis passing through the location of the observer,
and with its $y'$ and $z'$ axis perpendicular to the $x'$ axis.
If we know the coordinates $(x_i, y_i, z_i)$ of a mass element
in the old coordinate system (centered on the observer), we
can compute the coordinates $(x_i', y_i', z_i')$ in the new
coordinate system (centered on the void center).

A dimensionless measure of a void's elongation or compression along the
line of sight is
\begin{equation}
Q \equiv {3 \overline{(x')^2} \over \overline{(x')^2} +
\overline{(y')^2} + \overline{(z')^2} }  \ .
\end{equation}
The quantity $Q$ has some useful properties. Its denominator
is independent of the orientation of the coordinate system;
it's simply the mean square distance of all the elementary cells
from the void center. For a void of arbitrary shape, the mean
value of $Q$, averaged over all viewing angles, is $<\!Q\!> = 1$.
Thus, for a population of voids oriented randomly with respect
to the observer, we expect the average value of $Q$ to be 1.
A value $Q > 1$ indicates that a void is elongated along
the line of sight; a value $Q < 1$ indicates that a void
is compressed along the line of sight. For any void, in any
orientation, $0 \leq Q \leq 3$. As an example, consider a
triaxial ellipsoid with principal axes of length $a \geq b \geq c$.
The maximum value of $Q$ for this ellipsoid occurs when the
long axis is aligned with the line of sight from the observer
to the ellipsoid's center. In this case,
\begin{equation}
Q_{\rm max} = {3 a^2 \over a^2 + b^2 + c^2} \geq 1.
\end{equation}
The minimum value of $Q$ occurs when the short axis is aligned
with the line of sight. In this case,
\begin{equation}
Q_{\rm min} = {3 c^2 \over a^2 + b^2 +c^2} \leq 1.
\end{equation}

For the voids found by the AM algorithm,
the denominator of $Q$ can be written as
\begin{equation}
\overline{(x')^2} + \overline{(y')^2} + \overline{(z')^2} =
\overline{D^2} - \overline{D}^2 \ ,
\end{equation}
where
\begin{equation}
\overline{D}^2 = \overline{x}^2 + \overline{y}^2 + \overline{z}^2 
\end{equation}
and
\begin{equation}
\overline{D^2} = \overline{x^2} + \overline{y^2} + \overline{z^2} \ .
\end{equation}
The numerator can be written as
\begin{equation}
3 \overline{(x')^2} = {3 \over \overline{D}^2} \left[ - \overline{D}^4
+ \overline{x}^2 \overline{x^2} + \overline{y}^2 \overline{y^2}
+ \overline{z}^2 \overline{z^2} + 2 \overline{x} \, \overline{y} \, \overline{xy}
+ 2 \overline{y} \, \overline{z} \, \overline{yz} + 2 \overline{z} \, \overline{x}
 \, \overline{zx} \right] \ .
\end{equation}
To compute the distribution of $Q$ as a function of void volume
$V$, we first identify voids using the AM algorithm with a density
threshold $f=0.2$; to eliminate distortions caused by the artificial
boundary conditions at $z_{\rm max} = 0.12$, we consider only those
voids at redshifts $z < 0.11$. The mean, $\mu_Q$, and the standard
deviation, $\sigma_Q$, of $Q$ for the voids found in this way are
plotted in Figure 7, as a function of void volume $V$. The values
of $\mu_Q$ and $\sigma_Q$ are computed in bins containing 400
voids apiece. Thus, the expected error in the mean value of $Q$
for each bin is $\sigma_\mu = \sigma_Q / \sqrt{400} = 0.05 \sigma_Q$.
The value of $\mu_Q$ indicates whether voids are preferentially oriented
with respect to the line of sight; the value of $\sigma_Q$
is a measure of the intrinsic asphericity of the voids.

In the left column of Figure 7, the shape of voids is measured in real
space, where there should be no preferential distortions along
the line of sight. Indeed, $\mu_Q$ is not significantly
different from unity for voids of all sizes and at all
sampling densities: in all cases, $|\mu_Q - 1| < 2.5 \sigma_\mu$.
There is, however, a significant trend in the standard
deviation of $Q$. As voids get larger, $\sigma_Q$ gets smaller,
decreasing from $\sigma_Q \approx 0.34$ at $V = 5 \times 10^{-8}$
to $\sigma_Q \approx 0.27$ at $V > 10^{-6}$. This is a consequence
of the fact that large voids are more nearly spherical than small voids.

In the right column of Figure 7, the shape of voids is measured
in redshift space, where the distortions due to peculiar velocities
may cause systematic distortions along the line of sight. At
sufficiently high sampling density (the $L > L_*$ and $L > L_*/2$ samples),
large voids are significantly elongated along the line of
sight ($\mu_Q > 1$). In the $L > L_*$ sample, $\mu_Q > 1$ for
volumes $V > 4 \times 10^{-7} \sim 0.2 \overline{V}$. In the
$L > L_*/2$ sample, $\mu_Q > 1$ for volumes $V > 1.3 \times 10^{-7}
\sim 0.1 \overline{V}$; the greatest deviation of $\mu_Q$ from
unity is at $V = 1.25 \times 10^{-6} \sim \overline{V}$,
where $\mu_Q - 1 = 6 \sigma_\mu$. We conclude that the distortions
along the line of sight caused by peculiar velocities are
measurable by the AM algorithm only when the sampling density
is sufficiently high (corresponding to a limiting galaxy luminosity
of $\sim L_*$ or fainter). For our initial CDM power spectrum,
voids with $V \gtrsim \overline{V}$ show a significant tendency
to be elongated along the line of sight by peculiar velocities.

\section{Conclusion}

In examining the large scale structure of the universe,
studies of overdense regions (clusters and superclusters)
are usefully complemented by studies of underdense regions,
or voids. The properties of voids, such as their volumes
and shapes, depend on how voids are defined. Since the
literature contains many different void definitions and
void-finding algorithms, direct comparison of void properties
found in different studies is a risky business. 

The VPF, $P_0 (V)$, depends strongly on the number density
$\nu$ of galaxies in the survey. Although the UPF, $P_{0.2}
(V)$, depends less strongly on sampling density than the
VPF, it is still true that the characteristic void size
$V_{UPF}$ depends on the sampling density $\nu$, with
a higher density yielding a smaller characteristic void size.
Thus, in a flux-limited survey, where the
mean density of detected galaxies, $\nu (z)$, decreases with
redshift, using a threshold density $f \nu (z)$ will produce
a characteristic void size that increases with increasing
redshift. This will be an artifact of the decreasing sampling
density at large redshift, and does not reflect a change
in the underlying large scale structure. In determining
how the properties of voids depend on redshift, it is
more prudent to extract a volume-limited sample from the
data before applying a void statistic like the UPF or
a void-finding algorithm.

Of the void-finding algorithms outlined in this paper,
the ``sphere'' algorithm has the virtue of (relative)
simplicity. However, its restriction that all voids
must be spherical leads to an underestimate of void
size and does not permit us to measure the distortions
of voids caused by peculiar velocities. Adapting
the algorithm of \citet{am98}, we
were able to determine more accurately the sizes
of underdense voids. As in two-dimensional simulations
\citep{rm96}, the effect of peculiar velocities
is to increase the characteristic void size. The AM
algorithm also permits us to measure the elongation
of voids along the line of sight. In real space, it is
found that large voids are intrinsically more nearly spherical
than smaller voids. This can be regarded as a manifestation
of the tendency for large voids within a bubbly structure
to expand and become more nearly spherical at the
expense of their smaller neighbors \citep{rg91}.
In redshift space, large voids are seen, for the CDM
spectrum used in our simulations, to have a slight
but statistically significant tendency to be elongated
along the line of sight. Note, however, that the
void distortions can only be detected at a sufficiently
high sampling density. Future redshift surveys such
as that provided by the Sloan Digital Sky Survey will
provide sufficiently high galaxy densities and a large
enough number of voids to accurately measure the
peculiar velocity distortion of voids in the real universe.
When deeper redshift surveys are available, the (relatively
small) peculiar velocity distortions can be subtracted
out to reveal the cosmological distortions resulting
from the deceleration of the Hubble expansion.

\acknowledgments

This work was supported by a grant from the Ohio Supercomputer Center.
%Armen Ezekielian aided in the production of the three-dimensional
%visualizations.
ALM wishes to acknowledge support from the National
Science Foundation under grant number AST-0070702, the University
of Kansas General Research Fund and the National Center for
Supercomputing Applications.

\begin{figure}
\plottwo{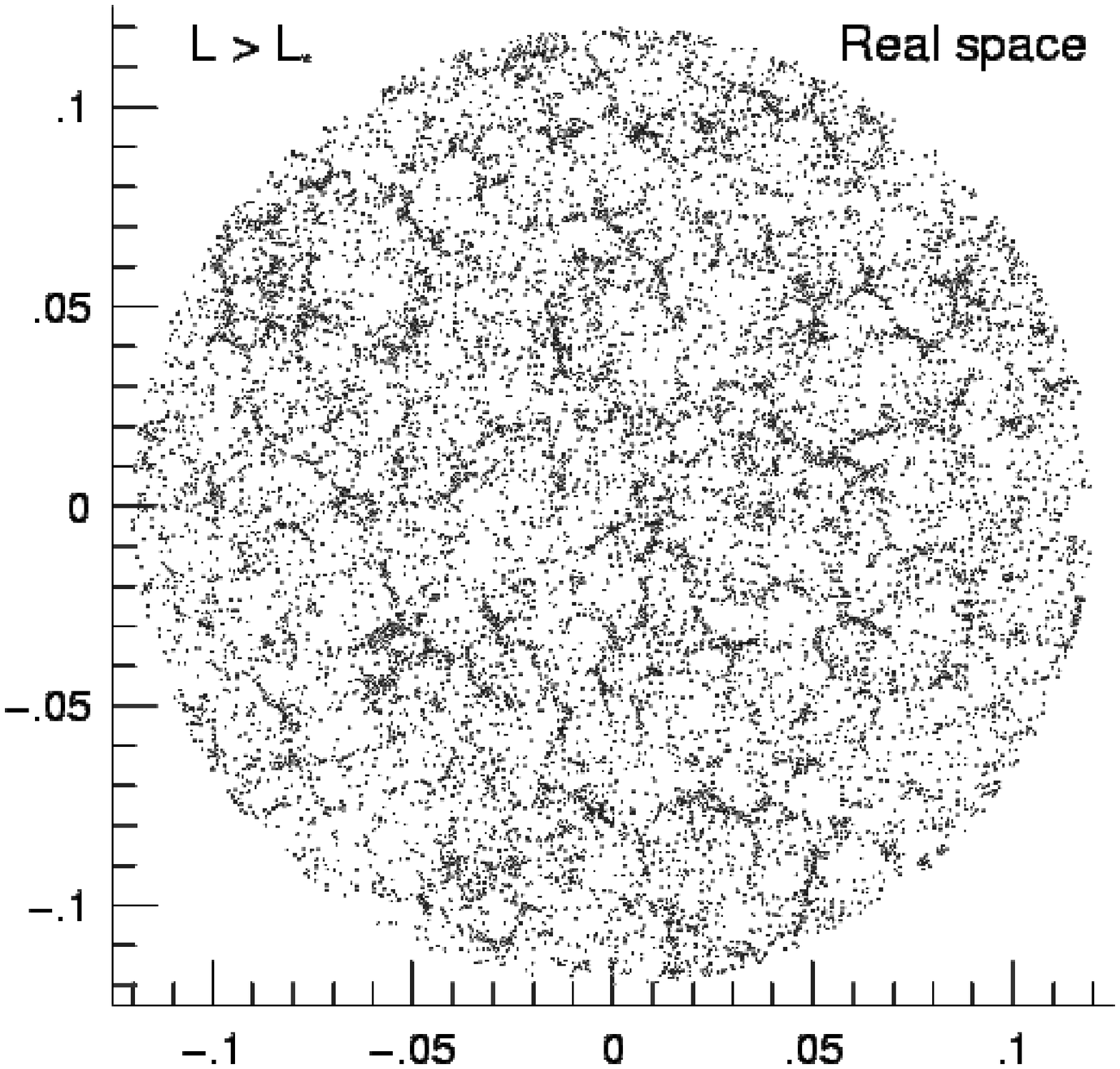}{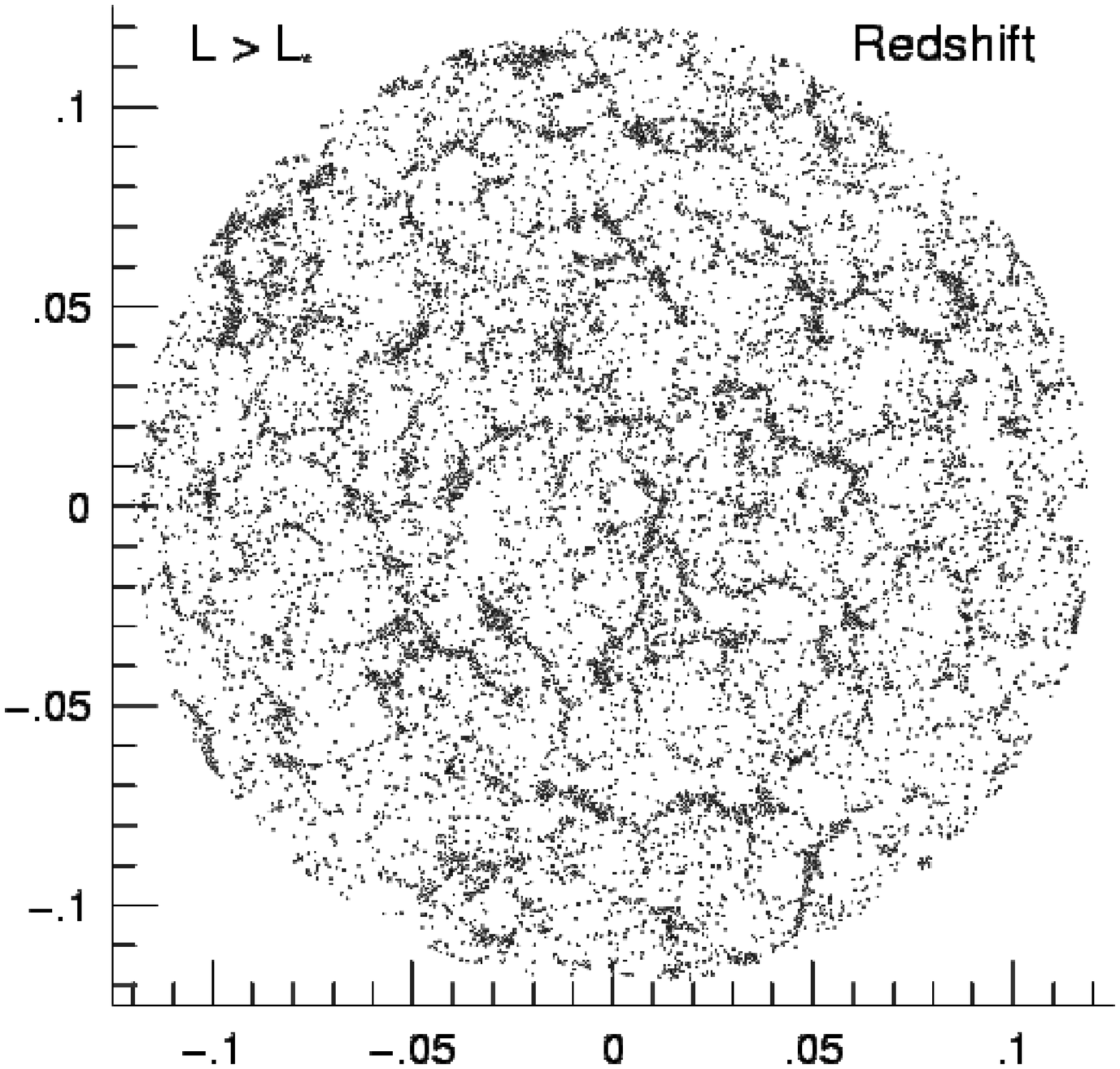}
\caption{(a) A slice of thickness $12 h^{-1} {\rm\,Mpc}$ through
the N-body simulation analyzed in this paper. The number density
of mass points is chosen to be comparable to the observed number density
of galaxies with $L > L_*$.  (b) A slice of the same thickness through
the same sample, but with peculiar velocities added. In each slice,
only galaxies with $z<0.12$ are included.}
\end{figure}

\begin{figure}
\plotone{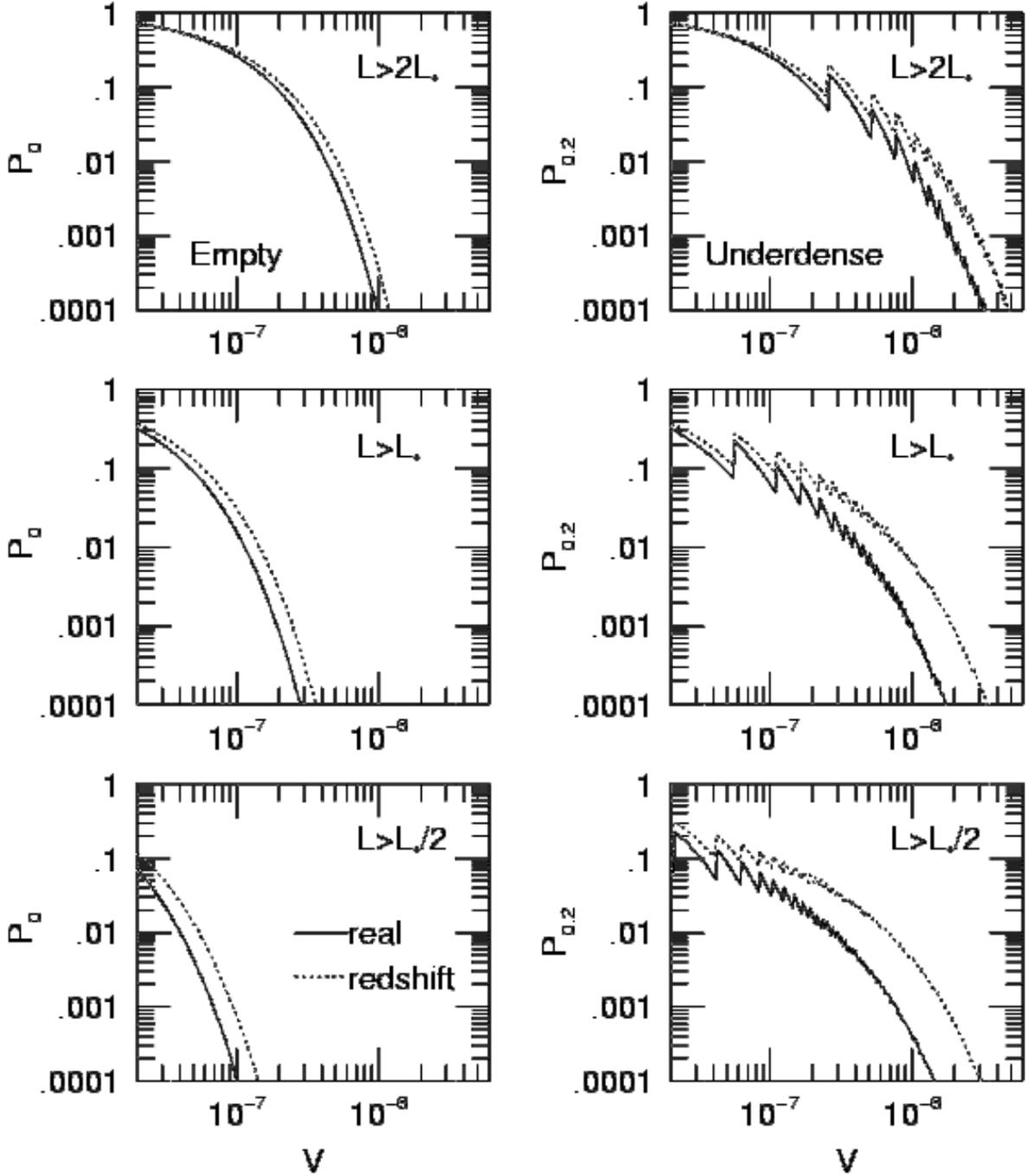}
\caption{The void probability function ({\it left column}) and underdense
probability function  ({\it right column}) for the CDM simulation.
In each column, the top panel represents the most sparsely sampled
subset of points (corresponding to $L > 2 L_*$),
and the bottom panel represents the most densely
sampled subset of points (corresponding to $L > L_*/2$). The volume
is given in dimensionless redshift units.}
\end{figure}

\begin{figure}
\plottwo{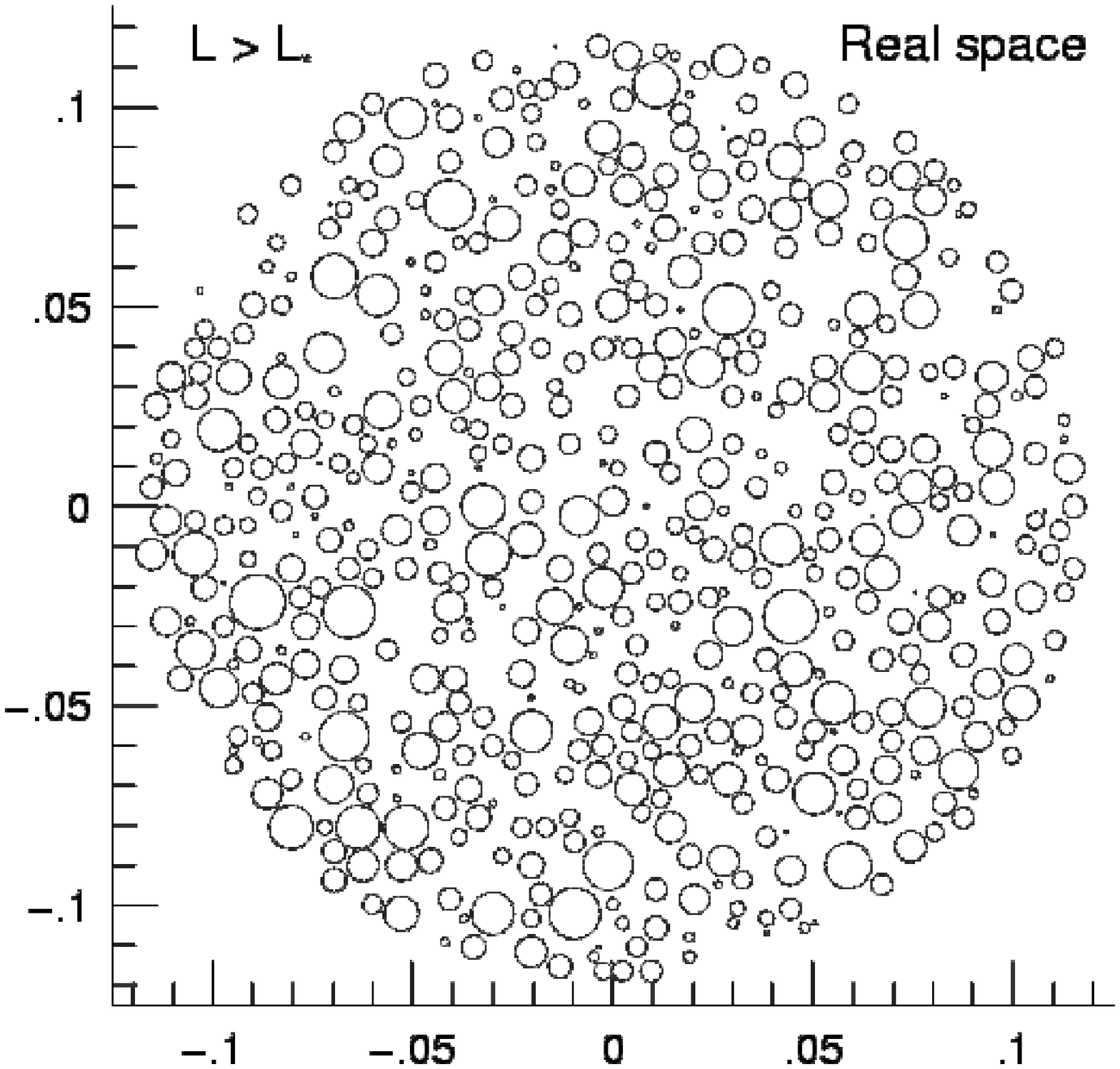}{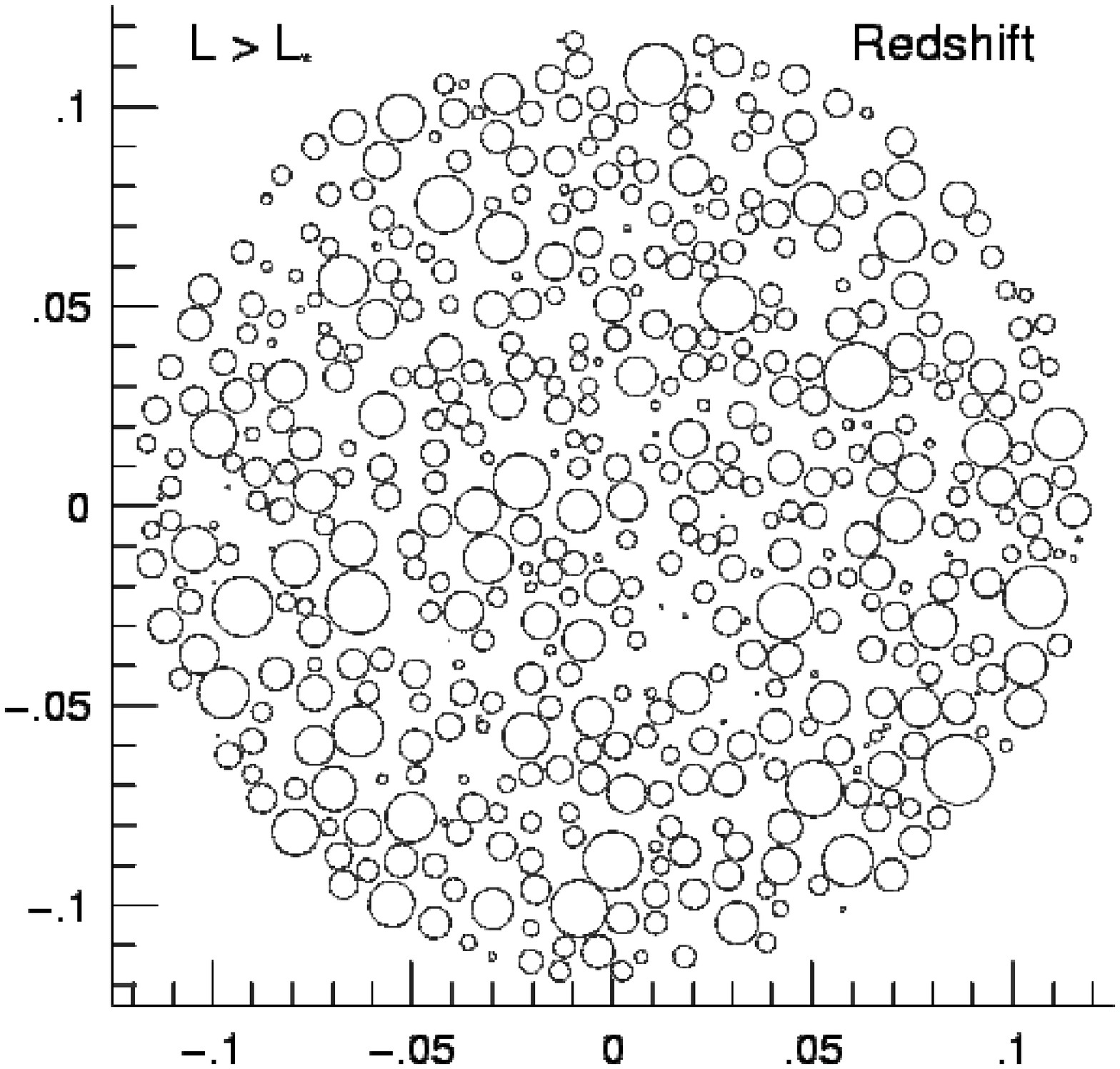}
\caption{(a) A slice through the spherical underdense voids found
in the $L > L_*$ sample; the density threshold chosen was
$f = 0.2$. The orientation of the slice is the same as that
shown in Figure 1a. (b) A slice through the spherical underdense
voids in the same sample, but with peculiar velocities added.}
\end{figure}

\begin{figure}
\plotone{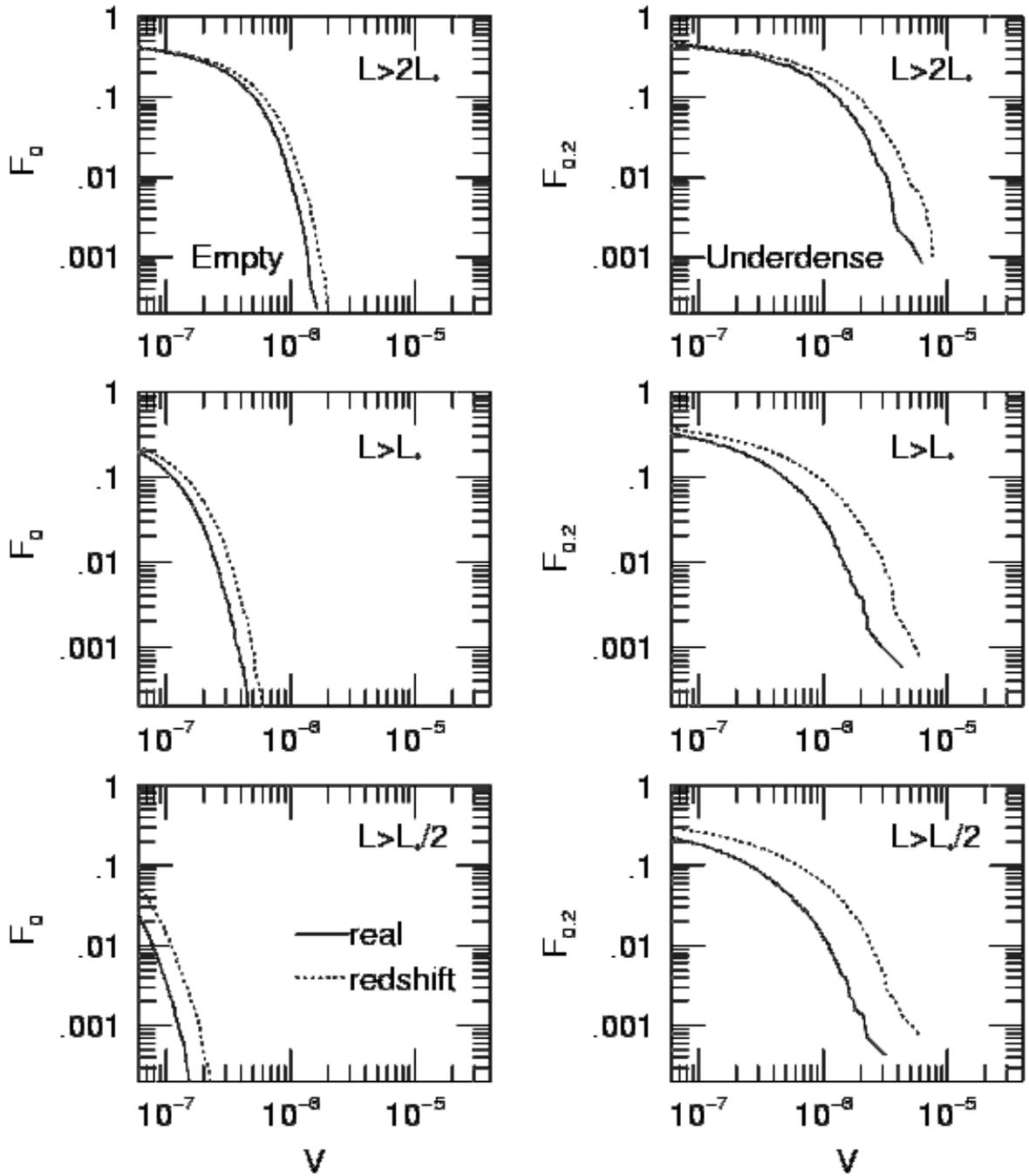}
\caption{Fraction of the total area of the simulation contained
in spherical voids of volume $V$ or greater. In the left column,
the density threshold is $f = 0$, corresponding to totally empty
voids. In the right column, the density threshold is $f = 0.2$.
In each panel, the solid line is the void fraction measured
in real space, and the dotted line is the void fraction measured
in redshift space.}
\end{figure}

\begin{figure}
\plottwo{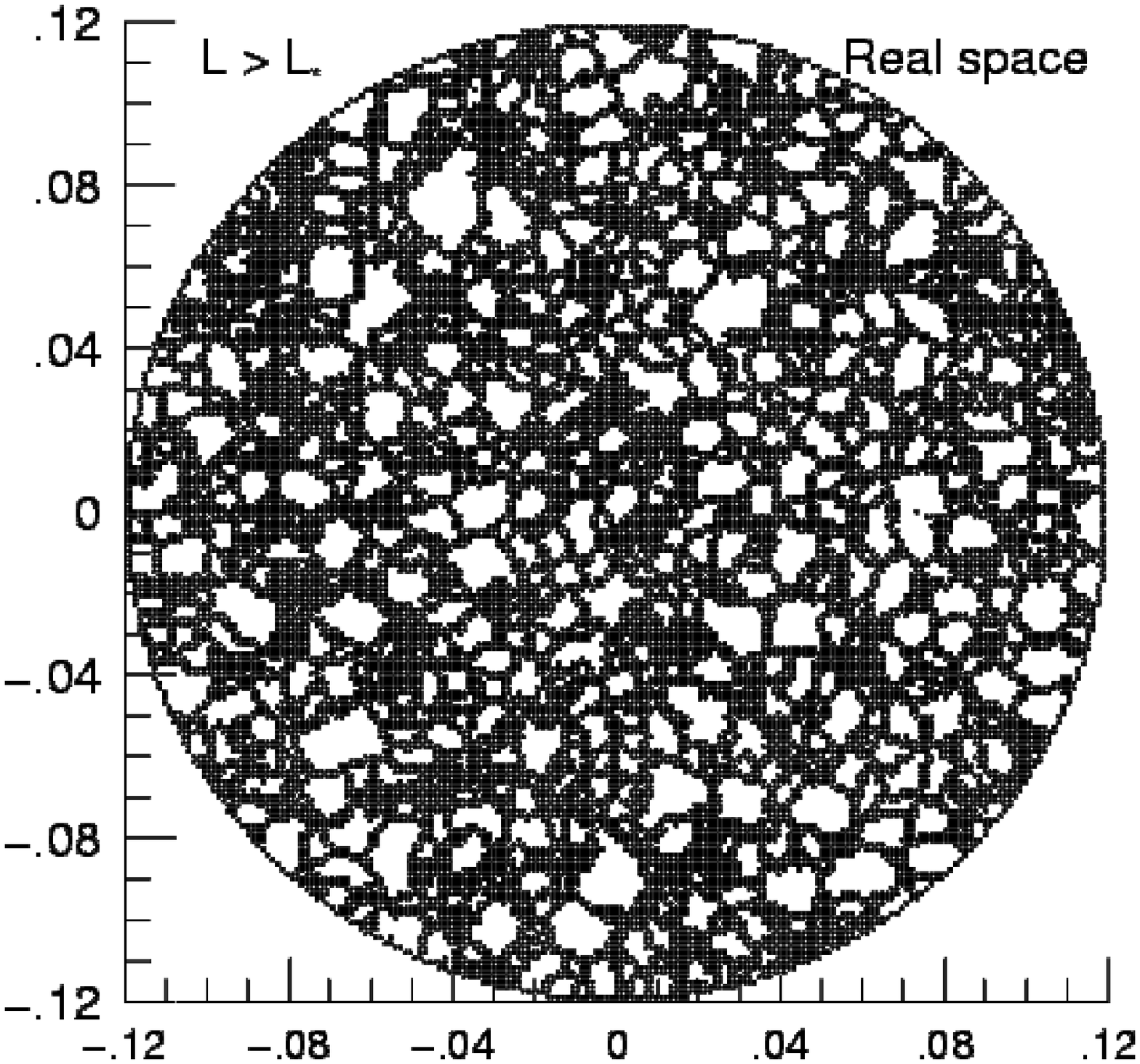}{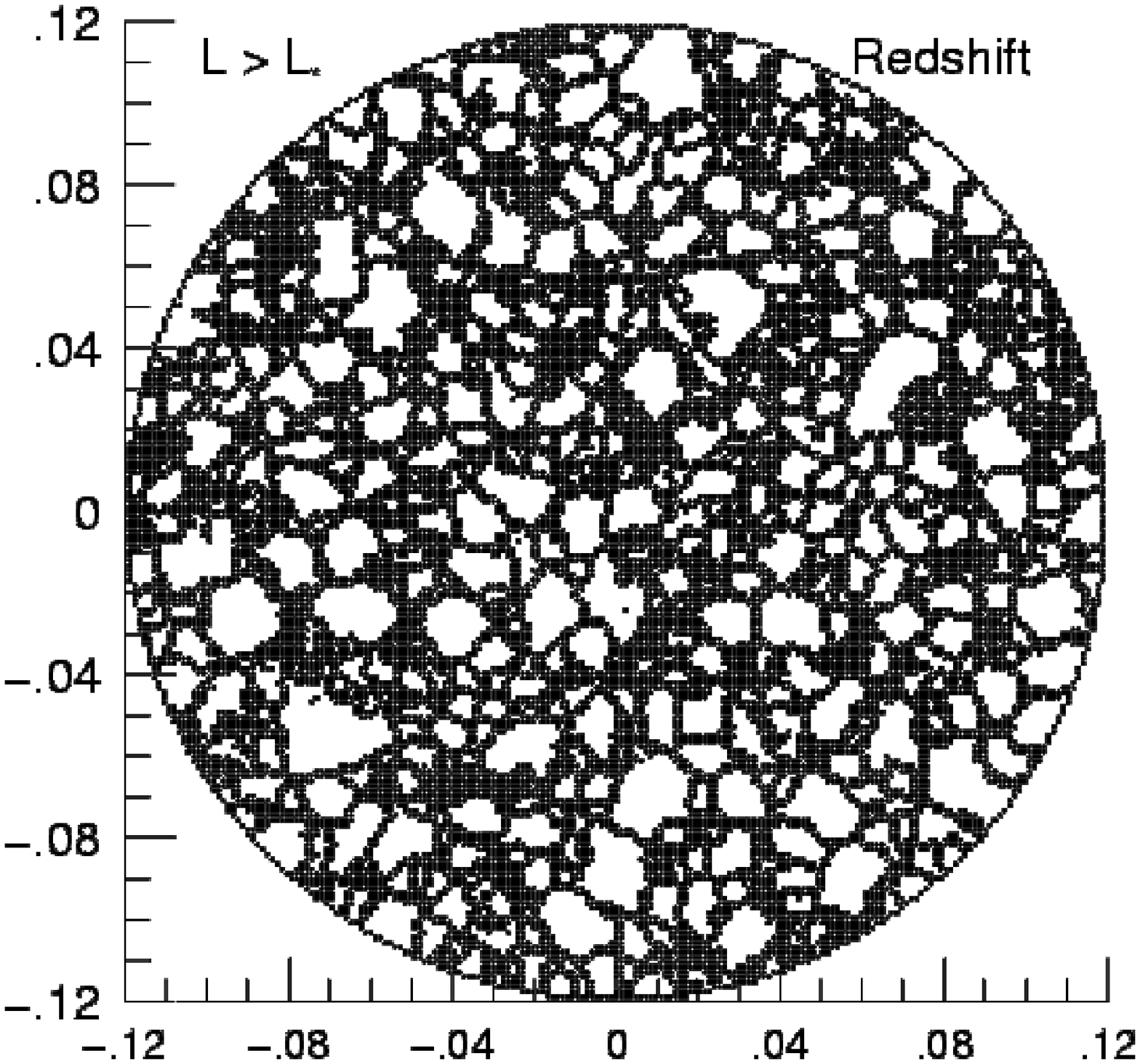}
\caption{(a) A slice through the underdense voids found in
the $L > L_*$ sample, using the AM algorithm; the density
threshold chosen was $f = 0.2$. The orientation of the
slice is the same as those in Figures 1a and 3a.
(b) A slice through the underdense AM voids in the same
sample, but with peculiar velocities added.}
\end{figure}

\begin{figure}
\plotone{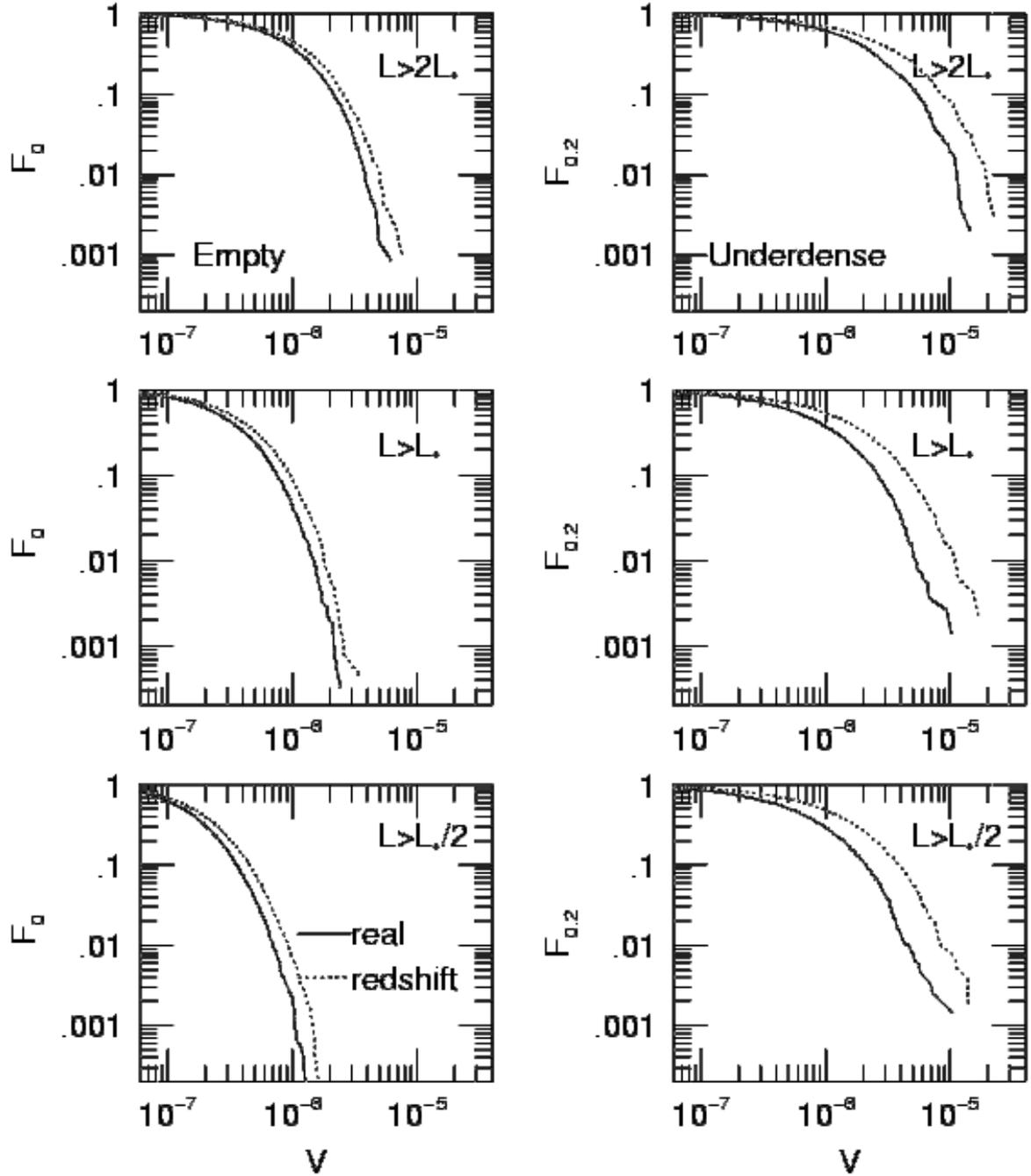}
\caption{Fraction of the total area of the simulation contained
in voids of volume $V$ or greater, as defined by the AM algorithm.
In the left column, the density threshold is $f=0$. In the
right column, the density threshold is $f=0.2$. In each panel,
the solid line is the void fraction measured in real space, and
the dotted line is the void fraction measured in redshift space.}
\end{figure}

\begin{figure}
\plotone{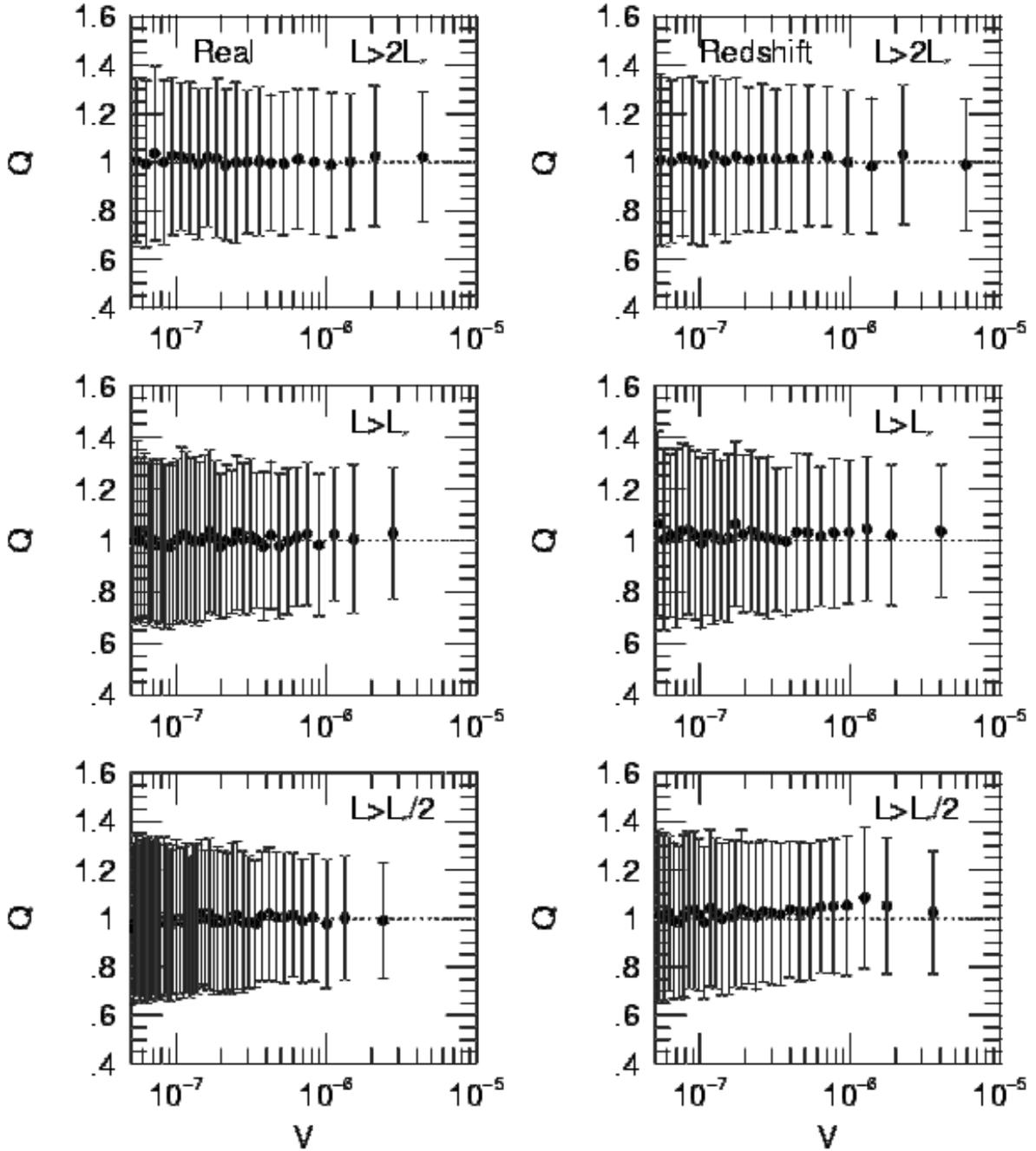}
\caption{The mean (filled circles) and standard deviation (error
bars) for the elongation statistic $Q$; each data point represents
a bin containing 400 voids. The left column gives the results
for voids measured in real space; the right column gives results
for voids measured in redshift space. In each case, the density
threshold is $f=0.2$.}
\end{figure}

\begin{table}
\caption{CHARACTERISTIC VOID SIZES}
\medskip
\medskip
\begin{tabular}{llcccc} 
& & \multicolumn{2}{c} {Spherical voids} &  \multicolumn{2}{c} {AM voids} \\
\multicolumn{2}{c}{Sample} & $\overline{V}$ & $V_{20}$ & $\overline{V}$ & $V_{20}$ \\
\tableline
$L > 2 L_*$ & (real) & 0.82E-6 & 0.67E-6 & 2.26E-6 & 3.49E-6 \\
& (redshift) & 1.17E-6 & 0.93E-6 & 3.57E-6 & 5.80E-6 \\
$L > L_*$   & (real) & 0.42E-6 & 0.21E-6 & 1.12E-6 & 1.75E-6 \\
& (redshift) & 0.71E-6 & 0.39E-6 & 1.95E-6 & 3.08E-6 \\
$L > L_*/2$ & (real) & 0.34E-6 & 0.085E-6 & 0.91E-6 & 1.39E-6 \\
& (redshift) & 0.64E-6 & 0.22E-6 & 1.63E-6 & 2.53E-6 \\

\end{tabular}
\end{table}

\end{document}